\documentstyle[mysprocl]{article}

\input{epsf}

\def\Journal#1#2#3#4{{#1} {\bf #2}, #3 (#4)}

\def\NPB{{\em Nucl. Phys.} B}
\def\PLB{{\em Phys. Lett.}  B}
\def\PRL{\em Phys. Rev. Lett.}
\def\PRD{{\em Phys. Rev.} D}
\def\ZPC{{\em Z. Phys.} C}
\def\EPJ{{\em Europ. Phys. J.} C}
\def\CPC{\em Comput. Phys. Commun.}

\def\be{\begin{equation}}
\def\ee{\end{equation}}
\def\bea{\begin{eqnarray}}
\def\eea{\end{eqnarray}}

\begin{document}

\title{Photon Structure in $\gamma p$ Interactions\footnote{To appear in the proceedings of the DIS98 workshop held in Brussels (Belgium), April 4-8, 1998.}\footnote{The inclusive jet photoproduction results presented at the workshop have been published \cite{incl}, and will not be covered in this paper}}

\author{J. H. VOSSEBELD}

\address{NIKHEF, P.O. Box 41882, 1009 DB Amsterdam, The Netherlands\\ 
E-mail: vossebeld@nikhef.nl}


\maketitle
\begin{center}
On behalf of the ZEUS collaboration.
\end{center}
\bigskip
\abstracts{Photoproduction of jets at 
HERA provides information on the partonic 
structure of the photon. We report on the latest dijet photoproduction 
results, for real photons 
and for photons at low virtualities, measured with the ZEUS detector.}

\section{Introduction}
In the $ep$-collider HERA\footnote{For the data presented in 
this paper 820 GeV protons were collided with 27.5 GeV positrons.} 
we can study, in addition to $ep$ Deep Inelastic Scattering, also 
$\gamma p$ scattering.
The photoproduction of jets is a process, well suited to 
test perturbative QCD and to study the partonic 
structure of the photon and the proton. In this paper we concentrate on
the structure of the photon.

In leading order we distinguish two classes of photoproduction processes: 
in the {\it direct} process the photon couples directly 
to a parton in the proton while in the {\it resolved} process a 
parton inside the photon couples to a parton in the proton. 
Kinematically the two classes can be separated using $x_\gamma$, 
the fractional momentum of the photon participating in the hard scatter.
For direct events $x_\gamma\sim1$ while for resolved events $x_\gamma<1$.
In resolved processes the partonic structure of the photon is probed.

We study events in which at least two hard jets are 
produced. Hard jets are associated with  high $P_T$ partons produced in the 
subprocess. The observation of hard jets thus ensures 
the presence of a hard scale. 
We present cross section measurements using two types of jetfinders. 
An iterative cone algorithm, according to the Snowmass convention \cite{snowmass}, and a 
$k_T$-clustering algorithm \cite{kt}.
For comparison of data with higher order QCD calculations the 
$k_T$-clustering algorithm is preferred. The ``Snowmass'' cone algorithm is not 
unambiguously defined \cite{rsep} and has numerical instabilities at 
higher orders \cite{coneprobl}.

\section{Dijet photoproduction}
Dijet photoproduction is particularly well suited to study the partonic structure of the 
photon, because
the reconstruction of two jets in the final state allows us to determine 
the kinematics of the subprocess. 
From the transverse energy ($E_T^{jet}$) and the pseudorapidity\footnote{$\eta^{jet}\equiv -ln(\tan{\frac{\theta^{jet}}{2}})$} ($\eta^{jet}$) of the jets 
one obtains the fractional momentum of the photon participating in 
the hard scatter:
\be
x_\gamma^{obs}=\left(E_{T\,1}^{jet}e^{-\eta_1^{jet}}+E_{T\,2}^{jet}e^{-\eta_2^{jet}}\right)/2yE_e
\label{eq:xg}
\ee
where $E_e$ is the beam energy of the incoming positron and $y$ is 
the fractional energy transfer from the positron 
to the proton in the proton rest frame.

The dijet cross section is measured as a function of the 
transverse energy and the pseudorapidity of the jets in the following kinematic range:
$-1<\eta_{1,2}^{jet}<2$, $E_{T\,1}^{jet}>14\,GeV$,  $E_{T\,2}^{jet}>11\,GeV$\footnote{The application of an asymmetric cut on the transverse energies of the two jets is required to ensure the stability of the NLO calculations \cite{symmcut}\label{etcutnote}.\vspace{-.2in}}, $0.20<y<0.85$ and $Q^2<1\,GeV^2$.
Jets are defined using a $k_T$-clustering algorithm.

From previous jet photoproduction analyses at lower transverse energies 
\cite{incl,mi}, we have 
learned that the cross sections  
show an excess over Monte Carlo predictions for forward jets, 
or correspondingly, for  low 
$x_\gamma^{obs}$ values. The data are well
described when a multiple interaction model is included in the Monte 
Carlo simulation.

\begin{figure}[ht]
\vspace{-.29in}
\epsfxsize = 2.85in
\epsfysize = 2.in
\epsffile{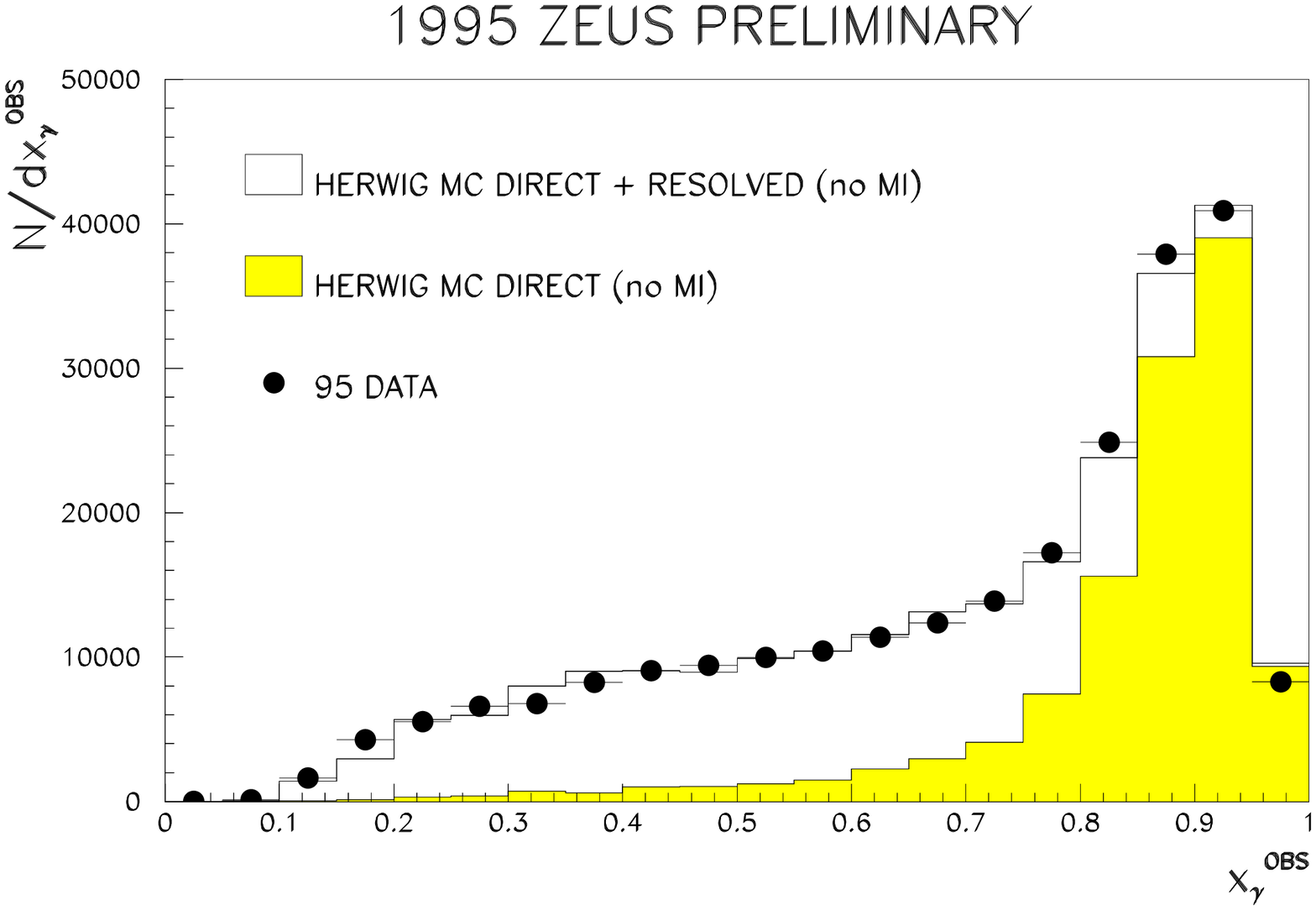}
\vspace{-1.8in}
\hspace{2.85in}
\epsfxsize = 2.in
\epsfysize = 1.7in
\epsffile{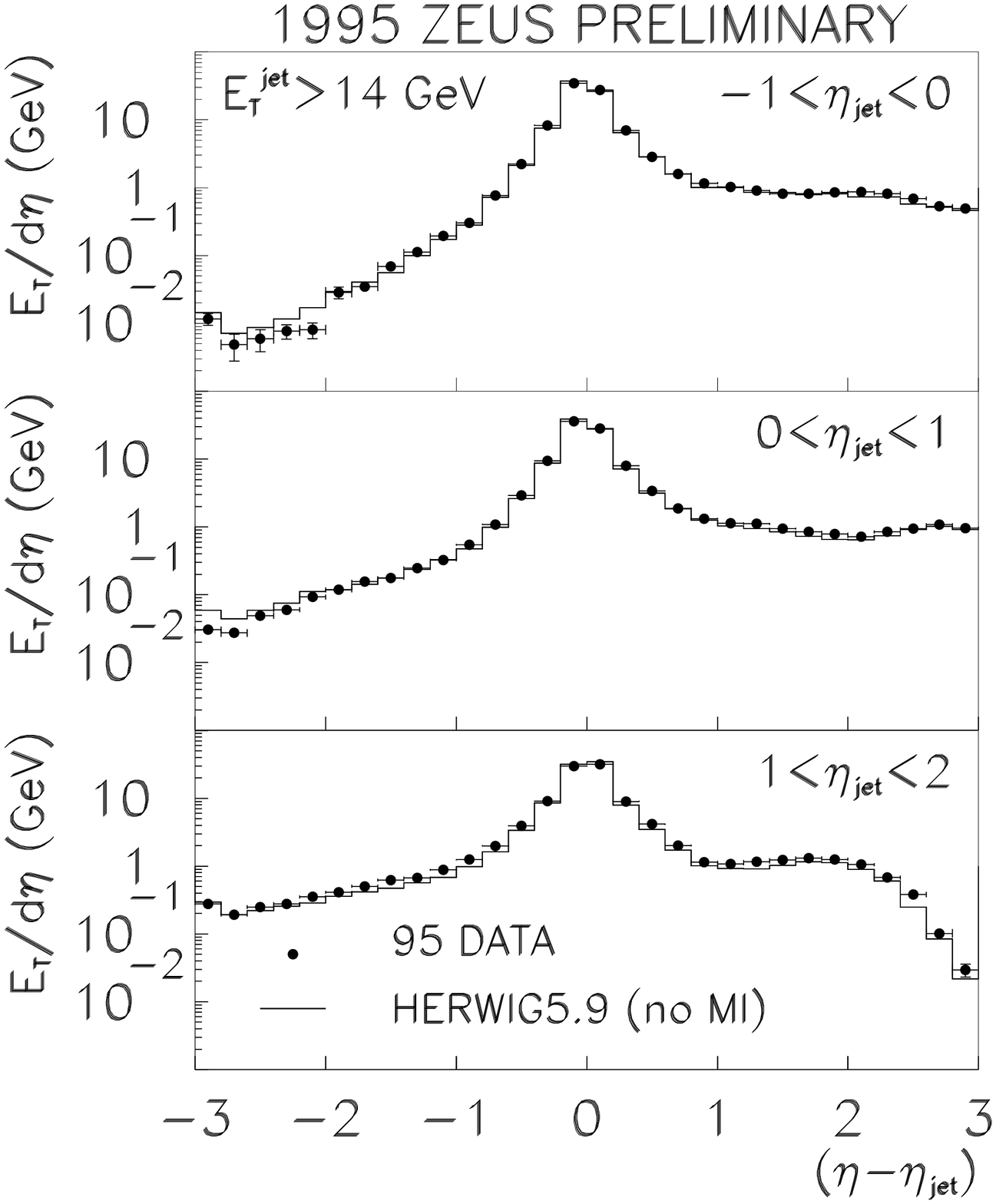}
\vspace{-.17in}
\caption{The uncorrected $x_\gamma^{obs}$ spectrum (left plot) and the transverse energy flow around jets (right plots), compared to Herwig 5.9. The transverse energy is plotted as a function  of the distance in pseudorapidity with respect to the jet (integrated over $\vert\Delta\phi\vert<1$). 
\label{fig:xgetflow}}
\vspace{-.09in}
\end{figure}

Requiring two hard jets reduces the 
sensitivity of the measured cross section to multiple interaction 
effects. 
As is shown in figure  \ref{fig:xgetflow} the measured spectra of 
$x_\gamma^{obs}$ and the transverse energy flow around jets are in excellent
agreement with predictions of the Herwig 5.9 \cite{hw} Monte Carlo, 
which has been used without including the multiple interaction option.

\begin{figure}[h]
\vspace{-.22in}
\epsfxsize = 2.45in
\epsfysize = 2.in
\epsffile{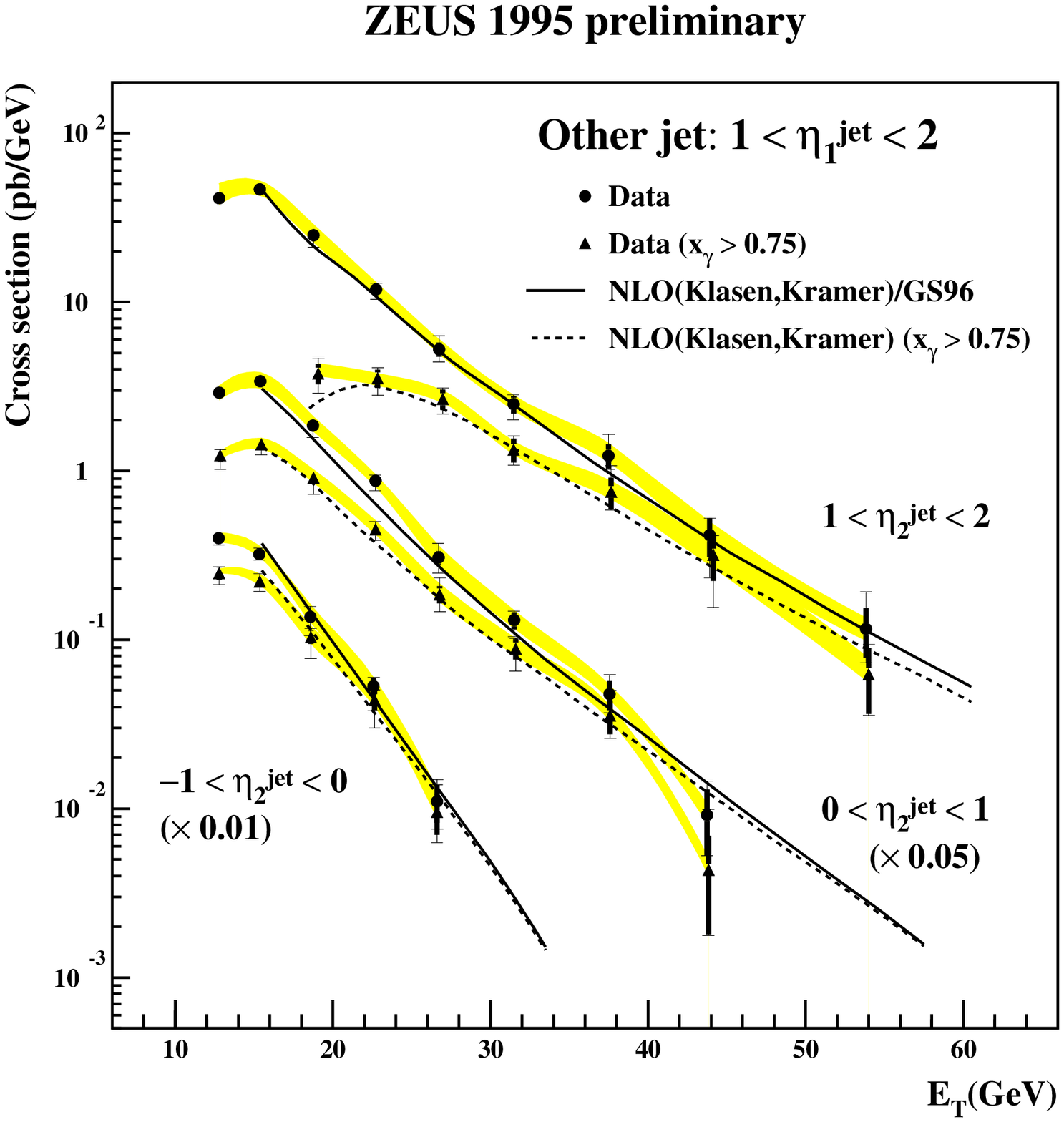}
\vspace{-2.in}
\hspace{2.4in}
\epsfxsize = 2.45in
\epsfysize = 2.in
\epsffile{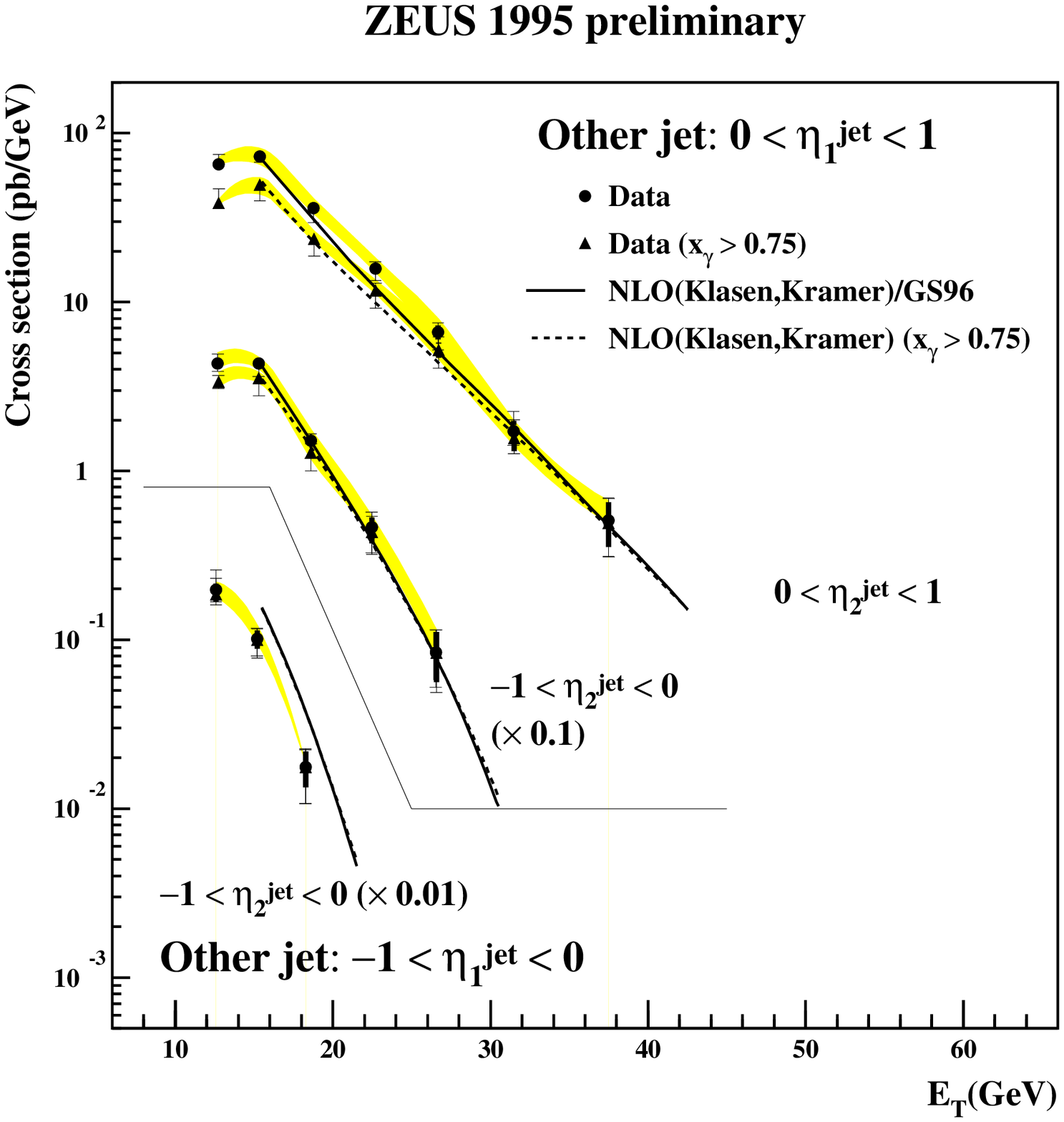}
\vspace{-.18in}
\caption{Inclusive dijet cross section as a function of $E_T^{jet}$ for various combinations of 
$\eta_1^{jet}$ and $\eta_2^{jet}$. The gray band indicates the (strongly correlated) uncertainty related to the energy scale.
\label{fig:dsigdet}}
\vspace{-.03in}
\epsfxsize = 5.in
\epsfysize = 1.2in
\epsffile{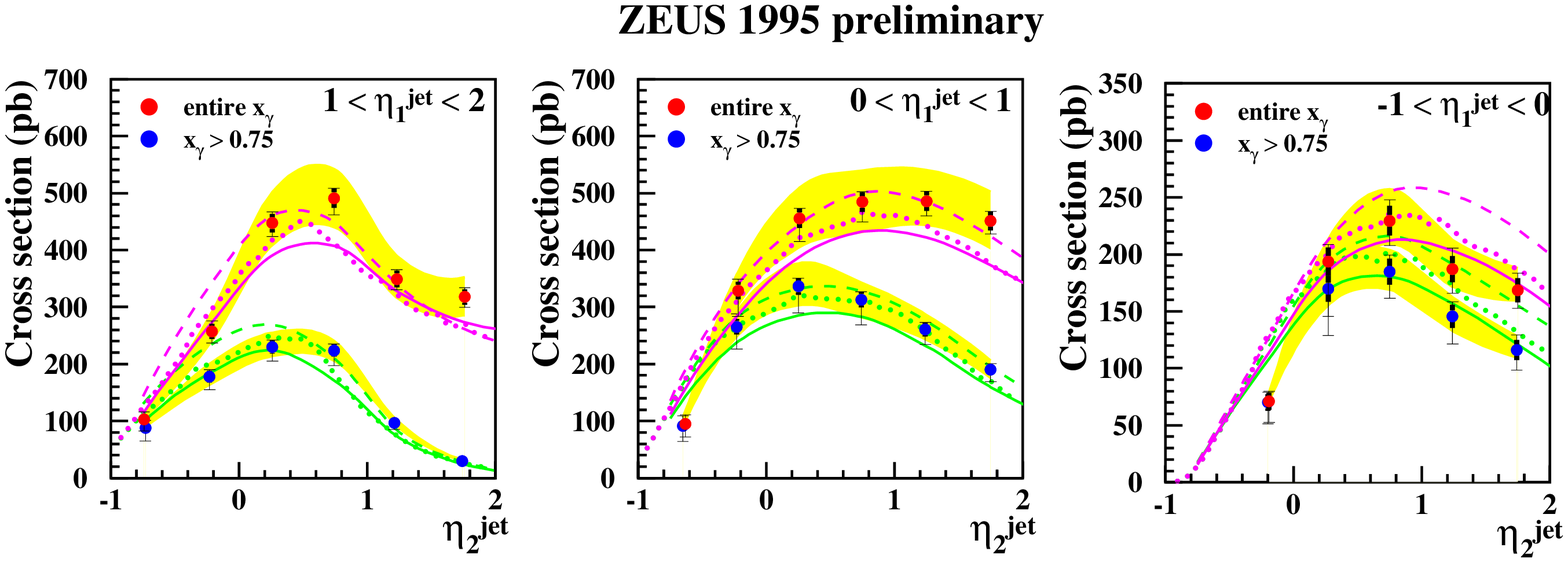}
\vspace{-.17in}
\caption{Inclusive dijet cross section as a function of $\eta_2^{jet}$ 
while $\eta_1^{jet}$ is fixed. The upper (lower) curves
correspond to the full (high) $x_\gamma^{obs}$ range. The full (dashed) curves are Kramer and Klasen NLO calculations using the GS96 (GRV) parameterization for the photon structure. The dotted curves are Harris and Owens calculations using GS96 for the photon structure.   
\label{fig:dsigdeta}}
\vspace{-.24in}
\end{figure}

Figure \ref{fig:dsigdet} and \ref{fig:dsigdeta} show the dijet cross section as a function of 
the transverse energy of the jets and of the pseudorapidity of the jets. Note that in both 
cases the cross section is symmetrized with respect to $\eta_1^{jet}$ and $\eta_2^{jet}$, therefore every event enters twice, once for every jet. 
Both for the full $x_\gamma^{obs}$-range and for the high $x_\gamma^{obs}$-range ($x_\gamma^{obs}>0.75$), the data agree well with NLO 
QCD predictions \cite{nlodijets}. 
We note that the 
differences related to the used photon parameterization \cite{photon} 
are of similar size as the uncertainty in the measurement.

\begin{figure}[b]
\vspace{-.25in}
\epsfxsize = 4.95in
\epsfysize = 1.2in
\epsffile{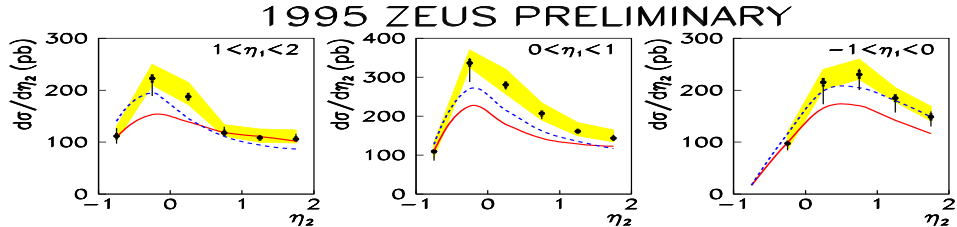}
\vspace{-.15in}
\caption{Inclusive dijet cross section as a function of $\eta_2^{jet}$ 
 while $\eta_1^{jet}$ is fixed, for high values of $y$. The full (dashed) curves
correspond to Klasen and Kramer NLO calculations using the GS96 (GRV) parameterization for 
the photon structure. \label{fig:highy}}
\vspace{-.13in}
\end{figure}

The dijet cross section is also measured as a function of the 
pseudorapidities of the jets in a restricted kinematic range 
($0.50<y<0.85$) to increase the sensitivity 
to the photon structure.
Effectively this restriction removes the highest $x_\gamma$ (direct) 
events in every bin.
In figure  \ref{fig:highy} the data are  
compared to NLO calculations using the same 
parameterizations of the photon structure as in figure \ref{fig:dsigdeta}. 
An increased sensitivity to the photon structure is observed in the NLO
calculations.
At central pseudorapidities, a significant discrepancy between the data
 and the QCD predictions is observed. 

Scale uncertainties \cite{nlodijets} and hadronization effects are expected to
be below $10\%$.

\section{Dijet production at low photon virtualities}
A subject of great interest is the production of dijets when the 
photon has a finite virtuality $Q^2$.
The contribution from resolved processes to the cross section is 
expected to be reduced for increasing values of $Q^2$. 

We have selected a dijet sample in the following kinematic range:
$-1.125<\eta_{1,2}^{jet}<1.875$, $E_{T\,1,2}^{jet}>6.5\,GeV$ and $0.2<y<0.55$.
Jets are defined with a cone algorithm.

The dijet cross section is determined in two ranges of $Q^2$: the  
untagged photoproduction region for which $Q^2\sim 0$ and 
the range $0.10<Q^2<0.70\,GeV^2$, which corresponds to events in which the 
scattered positron is detected in the beam pipe calorimeter 
\cite{bpc}.

\begin{figure}[h]
\vspace{-.15in}
\epsfxsize = 2.7in
\epsfysize = 2.3in
\epsffile{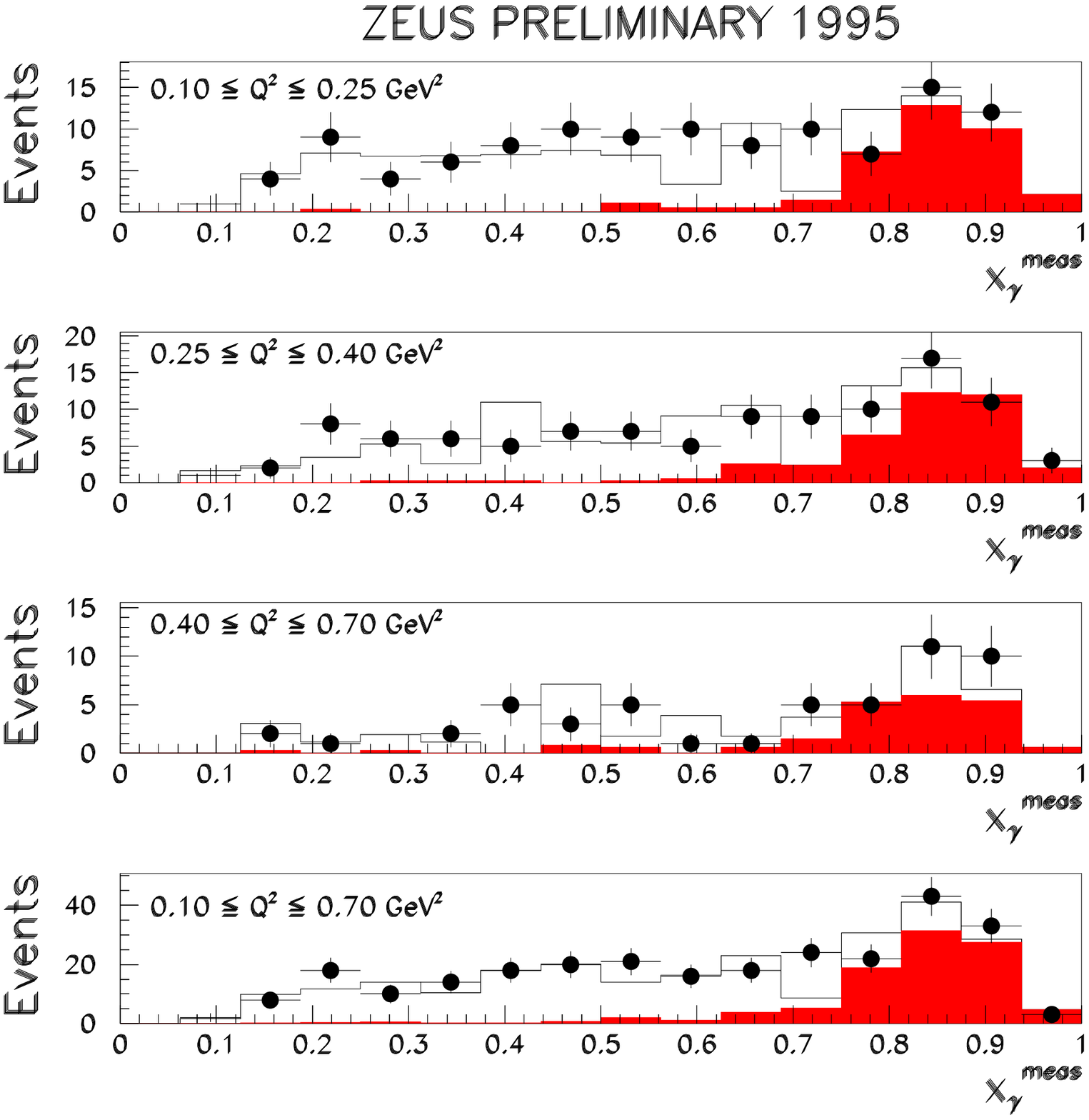}
\vspace{-2.3in}
\hspace{2.75in}
\epsfxsize = 2.1in
\epsfysize = 2.3in
\epsffile{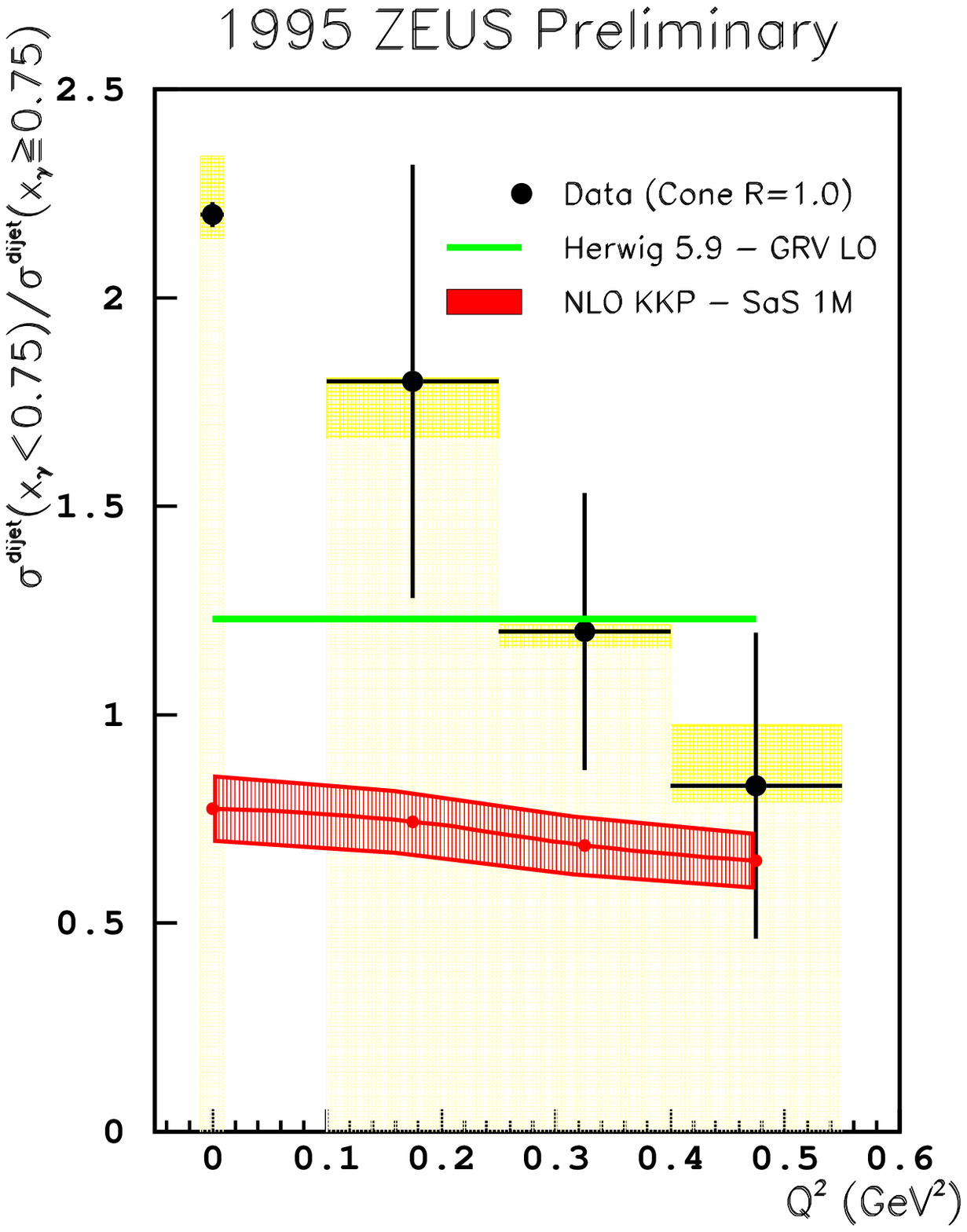}
\vspace{-.13in}
\caption{The uncorrected $x_\gamma^{obs}$ spectrum (left plot) in 
different bins of $Q^2$, compared to Herwig 5.9 (no multiple interactions and 
no $Q^2$ suppression). The hatched histogram shows the direct component of 
the Monte Carlo separately.
The right plot shows the ``resolved over direct'' ratio as a function of $Q^2$.
\label{fig:virtxgR}}
\vspace{-.1in}
\end{figure}

In figure \ref{fig:virtxgR} the $x_\gamma^{obs}$ distribution is 
compared to Herwig 5.9, in different bins of $Q^2$. 
For $Q^2$ values up to $0.70\,GeV^2$, both the resolved and 
the direct (dark histogram) component of the Monte Carlo are needed to describe the data.

Also in figure \ref{fig:virtxgR} the ``resolved over direct'' ratio of the cross section for $x_\gamma^{obs}<0.75$ over that for $x_\gamma^{obs}>0.75$ is plotted as a function of $Q^2$.
A decrease of this ratio with $Q^2$ is observed.
Herwig 5.9, which contains no mechanism for $Q^2$ suppression of the 
resolved component, can not reproduce this behavior. NLO calculations 
\cite{kkp} using the ($Q^2$ suppressed) SaS \cite{sas} parameterization 
for the photon, also fail to describe the data.
This might be related to the symmetric cut applied to the transverse energy 
of the two jets (see footnote \ref{etcutnote}).

\section{Summary}
Photoproduction of jets at HERA offers a 
unique source of information on the structure of the photon.
The dijet photoproduction measurements presented in this paper show a
discrepancy between data and NLO QCD predictions for high energy real photons.
For virtual photons we have observed  a decrease of the 
``resolved over direct'' ratio as a function of $Q^2$ 
(up to $Q^2=0.70\,GeV^2$).

\end{document}